# Beyond the Einstein–Bohr Debate: Cognitive Complementarity and the Emergence of Quantum Intuition

*Lalit Kumar Shukla[1]*

*1. Faculty of Physical Sciences, Shri Ramswaroop Memorial University Uttar Pradesh, India*
*Email: lalitshukla.phy@srmu.ac.in*

**Abstract:** Recent high-precision experimental confirmations of quantum complementarity have revitalized foundational debates about measurement, description, and realism. This article argues that complementarity is most productively interpreted as an epistemic principle—constraining what can be simultaneously accessed and represented—rather than as an ontological claim about quantum reality. Reexamining the Einstein–Bohr debate through this lens reveals a persistent tension between descriptive completeness and contextual meaning, a tension experiments clarify but do not dissolve. Building on this analysis, we introduce cognitive complementarity as a structural principle governing reasoning under non-classical uncertainty, where mutually constraining representations cannot be jointly optimized. Within this framework, we propose quantum intuition as a testable cognitive capacity: the ability to sustain representational plurality, regulate commitment timing, and resolve perspective-incompatibilities in a context-sensitive manner. Formulated as a naturalistic construct grounded in shared informational constraints, quantum intuition offers a principled bridge between quantum measurement theory and cognition. This work reframes the historical debate, extends epistemic lessons from quantum foundations into cognitive science, and outlines empirical pathways for studying decision-making in contexts of irreducible uncertainty.

## 1. Introduction

Quantum complementarity has long occupied a central position in the foundations of quantum mechanics, articulating the limits within which physical properties may be meaningfully ascribed to quantum systems. From its earliest formulations, complementarity emphasized that the manifestation of quantum phenomena depends crucially on the experimental arrangement, and that certain descriptive frameworks—while individually well-defined—cannot be simultaneously applied without loss of coherence (Bohr, 1928). Although historically expressed in qualitative terms, complementarity has since been given precise operational meaning through developments in quantum optics, information theory, and the theory of open quantum systems (Englert, 1996; Zurek, 2003).

Recent advances in experimental control have enabled realizations of interferometric and scattering setups that closely approximate idealized versions of classic thought experiments associated with the Einstein-Bohr debate (Durr et al., 1998; Zhang et al., 2025). Employing single atoms, single photons, and finely tunable measurement interactions, these experiments demonstrate with high precision the quantitative trade-off between path distinguishability and fringe visibility. Notably, entanglement-enabled delayed-choice experiments have further probed the boundaries of complementarity, showing that the wave-like or particle-like behavior of a quantum system can be decided after it has already entered the interferometer (Kaiser et al., 2012). Similarly, quantum delayed-choice experiments with single photons reinforce the contextual nature of quantum phenomena (Peruzzo et al., 2012). Reviews of such delayed-choice gedanken experiments confirm that complementarity cannot be circumvented by delaying measurement decisions, underscoring the fundamental role of context in defining quantum properties (Ma et al., 2016). These results provide strong empirical support for modern formulations of complementarity, in which the extraction of which-path information necessarily accompanies a reduction in interference contrast. Importantly, these outcomes remain fully consistent with standard quantum theory and with information-theoretic descriptions of measurement-induced decoherence (Zurek, 2003; Schlosshauer, 2007).

This renewed experimental attention to complementarity has naturally prompted renewed conceptual discussion. In particular, questions have resurfaced concerning the extent to which operational limits on measurement constrain our understanding of quantum reality itself. While complementarity unambiguously restricts the simultaneous accessibility of certain observables, it remains an open interpretive question whether such restrictions should be read as limits on knowledge, on description, or on the existence of underlying physical properties (Fine, 1986; Howard, 1994). Addressing this distinction requires careful separation of epistemic claims—concerning what can be known or measured—from ontological claims—concerning what may exist independently of measurement.

Historically, this distinction lay at the heart of the Einstein-Bohr debate. Einstein's reservations about quantum mechanics were not directed at its empirical adequacy, but at what he perceived as its incompleteness as a description of physical reality (Einstein et al., 1935). Bohr, in contrast, emphasized the contextual conditions under which physical quantities acquire meaning, arguing that quantum mechanics exhausts what can be said within those conditions (Bohr, 1949). Subsequent experimental confirmations of quantum predictions have repeatedly favored Bohr's operational conclusions, yet they have not eliminated the conceptual space for questions about descriptive completeness or realism. Modern experiments, despite their sophistication, inherit this same limitation: they test the structure of quantum predictions under controlled conditions, but they do not by themselves resolve interpretive questions about ontology.

Contemporary formulations of complementarity clarify this situation by embedding it within an information-theoretic framework. The trade-off between path distinguishability and fringe visibility can be expressed quantitatively, and its origin can be traced to the entanglement of the system with measurement degrees of freedom and the environment (Englert, 1996; Coles et al., 2014). From this perspective, the suppression of interference arises from the dispersal of phase information into correlations that are no longer locally accessible. Decoherence theory accounts for the emergence of classical-like outcomes without requiring any modification of unitary quantum dynamics, while remaining neutral with respect to deeper ontological interpretations (Zurek, 2003). Consequently, complementarity is most robustly understood as a constraint on representation and information access, rather than as a statement about the nonexistence of unmeasured properties.

The aim of the present work is to develop this epistemic reading of complementarity and to explore its implications beyond the immediate domain of quantum measurement. While the physical content of complementarity is well established, its broader significance for how agents reason under conditions of non-classical uncertainty has received comparatively little systematic attention. The limits imposed by complementarity are not merely technical features of experimental design; they exemplify a more general situation in which no single descriptive framework suffices to capture all relevant aspects of a system simultaneously. This observation motivates an inquiry into whether similar structural constraints operate in cognitive contexts where information is contextual, non-commuting, or irreducibly incomplete (Busemeyer & Bruza, 2012; Pothos & Busemeyer, 2009).

Accordingly, this article proceeds in two stages. First, it offers a careful reassessment of complementarity and the Einstein-Bohr debate, emphasizing the distinction between operational constraints and ontological claims. Second, it introduces a conceptual extension of complementarity into the cognitive domain, proposing that certain features of human reasoning under uncertainty mirror the representational limits encountered in quantum measurement. This extension is developed without assuming that cognition is physically quantum, and without invoking speculative claims about consciousness or measurement. Instead, the analysis focuses on structural parallels at the level of information, representation, and decision commitment.

By maintaining a clear separation between established physical results and proposed cognitive interpretations, we aim to contribute both to ongoing discussions in quantum foundations and to emerging approaches in the study of reasoning under uncertainty. The introduction of cognitive complementarity and quantum intuition, developed in later sections, is intended not as a reinterpretation of quantum mechanics, but as an application of its epistemic lessons to the theory of cognition. In this way, the work seeks to clarify what recent experiments do—and do not—imply about reality, while opening a principled avenue for interdisciplinary inquiry grounded in rigor rather than metaphor.

To clarify the conceptual architecture of this work, Table 1 distinguishes the four key constructs developed herein. While quantum complementarity provides the physical-epistemic template, cognitive complementarity extends this logic to mental representation without assuming quantum physicality. Quantum intuition, in turn, denotes the cognitive capacity to operate within such constraints—a capacity open to empirical study without recourse to quantum metaphysics.

Table 1: Conceptual Distinctions in This Framework

| Concept | Domain | Nature | Role in This Work |
|---|---|---|---|
| **Quantum Complementarity** | Quantum Physics | Physical, informational | Epistemic constraint on simultaneous access to incompatible observables (e.g., path vs. interference). |
| **Quantum Cognition** | Cognitive Science | Formal, probabilistic | Uses quantum mathematical structures to model cognitive phenomena (e.g., order effects, contextuality). |
| **Cognitive Complementarity** | Cognitive Science | Epistemic, structural | A principle governing reasoning where certain representations cannot be jointly optimized without loss. |
| **Quantum Intuition** | Cognitive Science | Capacity, regulative | The cognitive ability to manage epistemic superposition and regulate commitment under uncertainty. |

## 2. Quantum Complementarity: From Principle to Epistemic Constraint

The modern understanding of quantum complementarity has evolved from Bohr's qualitative principle into a precise, quantitative framework grounded in interferometry, information theory, and decoherence. This transition clarifies not only complementarity's operational content but also supports interpreting it as a constraint on knowledge and representation, rather than a statement about reality itself. Below, we trace this development—from the basic two-path interferometer to information-theoretic and decoherence-based accounts—and argue that complementarity is most coherently understood as an epistemic principle governing the simultaneous accessibility of incompatible descriptions.

### 2.1. From qualitative principle to quantitative formulation

Originally, quantum complementarity was formulated in qualitative terms, emphasizing the mutual exclusivity of certain experimental arrangements and the contextual nature of physical descriptions (Bohr, 1928). While conceptually influential, this formulation left open questions about the precise mechanisms by which complementary behaviors exclude one another, and the extent to which this exclusion admits quantitative characterization. Developments in quantum optics and interferometry have since addressed these questions by translating complementarity into experimentally measurable quantities, thereby clarifying its operational content (Greenberger & Yasin, 1988; Englert, 1996). This translation includes revisiting Einstein's recoiling-slit gedankenexperiment, which has been shown to adhere quantitatively to the same complementarity constraints governing modern interferometric setups.

The paradigmatic setting for this development is the two-path interferometer, where interference phenomena depend on the coherent superposition of amplitudes associated with distinct paths. In such systems, interference fringe observation is not an all-or-nothing property, but varies continuously. Similarly, information about which path a quantum system has taken need not be completely absent or fully available. This recognition led to formulations of complementarity that express the trade-off between these two aspects in precise, quantitative terms.

**2.2. Path distinguishability and fringe visibility**

Two quantities play a central role in modern discussions of complementarity: path distinguishability and fringe visibility. Path distinguishability, commonly denoted by $D$, characterizes the extent to which the two paths of an interferometer can be operationally distinguished, either through direct measurement or through correlations with auxiliary degrees of freedom. Fringe visibility, denoted by $V$, quantifies the contrast of the interference pattern and is defined as

$$V = \frac{I_{Max} - I_{Min}}{I_{Max} + I_{Min}}$$

where $I_{Max}$ and $I_{Min}$ represent the maximum and minimum detected intensities, respectively.

In ideal interference experiments, maximal fringe visibility ($V = 1$) corresponds to complete coherence between the paths and the absence of any which-path information. Conversely, perfect path distinguishability ($D = 1$) implies that the paths can be unambiguously identified, in which case interference fringes vanish ($V = 0$). Between these extremes lies a continuous range of intermediate regimes, reflecting partial coherence and partial path information.

The relationship between these quantities is captured by quantitative complementarity relations, most notably the inequality

$$V^2 + D^2 \leq 1$$

which holds under broad and experimentally relevant conditions. This relation demonstrates that path distinguishability and fringe visibility are not independent properties, but are constrained by a fundamental trade-off. This trade-off can be further understood as a duality between quantum coherence and path distinguishability, formalized in modern treatments of wave-particle complementarity (Bera et al., 2015). Importantly, the inequality does not privilege either quantity; rather, it delineates the space of jointly accessible values, emphasizing that increased access to path information necessarily limits the visibility of interference.

**2.3. Information, decoherence, and the loss of interference**

The quantitative formulation of complementarity admits a natural interpretation in information-theoretic terms. Path distinguishability can be understood as a measure of the information available about the system's trajectory through the interferometer, while fringe visibility reflects the degree to which phase information remains accessible in the observed interference pattern. The complementarity relation thus expresses a constraint on the simultaneous accessibility of different types of information encoded in the quantum state (Coles et al., 2014).

From this perspective, complementarity does not arise from a mysterious disturbance induced by measurement, but from the redistribution of information through correlations. When a quantum system becomes correlated with a which-path detector or with environmental degrees of freedom, information about its path is transferred to those systems. This transfer renders phase information locally inaccessible, even though it remains encoded in the global quantum state. The loss of interference is therefore a consequence of information becoming delocalized rather than destroyed. This process has been directly observed in experiments with single-atom wave packets, where coherent and incoherent scattering leads to a measurable loss of interference visibility, demonstrating decoherence in a controlled, single-particle context (Fedoseev et al., 2025).

Decoherence theory provides a dynamical account of how this trade-off emerges in realistic physical settings (Zurek, 2003). When a quantum system interacts with an environment, the system's reduced density matrix typically loses coherence in a preferred basis determined by the interaction. In interferometric contexts, this process suppresses the off-diagonal terms responsible for interference, effectively selecting a mixture of path states. This suppression has been directly observed in matter-wave interferometry experiments, where collisional decoherence causes a measurable loss of fringe visibility (Hornberger et al., 2003). Crucially, decoherence operates within standard unitary quantum mechanics and does not invoke any modification of the Schrödinger equation. The apparent transition from coherent superposition to classical alternatives arises from tracing over environmental degrees of freedom, which renders certain correlations inaccessible to local observers. Decoherence therefore explains the practical irreversibility of interference loss while remaining agnostic about deeper ontological questions, such as whether the global quantum state retains coherence at a fundamental level.

### 2.4. Complementarity as an epistemic principle

Modern formulations of complementarity, supported by information-theoretic analysis and decoherence theory, precisely delineate the limits of what can be operationally accessed in quantum experiments. These limits concern the simultaneous availability of certain descriptive features—such as path information and interference contrast—given specific experimental conditions. Crucially, they do not by themselves specify what exists independently of those conditions.

This distinction is essential for interpreting both historical debates and contemporary experiments. Complementarity constrains the structure of possible measurement outcomes and the representations that can be consistently employed, but it does not mandate a particular stance on the underlying ontology of quantum systems. As such, complementarity functions most robustly as an epistemic principle: it governs the conditions under which information can be extracted and meaningfully ascribed, while leaving questions of realism and completeness logically open (Howard, 1994; Bub, 1997).

In the following section, we examine this distinction in the context of the Einstein–Bohr debate. By reassessing Einstein's objections and Bohr's responses in light of the modern understanding of complementarity, we aim to clarify what is—and is not—settled by experimental confirmations of quantum predictions.

## 3. Reassessing the Einstein–Bohr Debate

The modern, epistemic reading of complementarity provides a clarifying lens through which to revisit the historic dialogue between Einstein and Bohr. Rather than framing their exchange as a conflict over the correctness of quantum mechanics, this perspective reveals a deeper disagreement about the aims and completeness of physical description. In this section, we examine Einstein's incompleteness objection, Bohr's contextual reply, and what modern experiments actually demonstrate about their competing views.

### 3.1. Einstein's incompleteness: A realist critique of quantum description

The historical debate between Albert Einstein and Niels Bohr is frequently summarized as a disagreement over the correctness of quantum mechanics. Such summaries obscure a crucial point: Einstein did not dispute the empirical success of the theory. On the contrary, he repeatedly acknowledged its extraordinary predictive power. His concern lay instead with what he regarded as the incompleteness of the quantum description of physical reality (Einstein et al., 1935).

Einstein's objection was rooted in a realist intuition according to which a physical theory should provide a description of individual systems that is independent of the act of measurement. In his view, a theory that could only assign probabilities to measurement outcomes, without specifying underlying physical properties, failed to meet this criterion of completeness. This position did not require adherence to classical determinism; rather, it reflected a demand that physical quantities correspond to elements of reality, even when not directly observed (Fine, 1986). From this perspective, complementarity appeared problematic not because it contradicted experiment, but because it seemed to elevate limitations of measurement to statements about reality itself.

### 3.2. Bohr's contextualism: Meaning as conditioned by experimental arrangement

Bohr's reply to Einstein did not consist in denying the existence of an underlying reality. Instead, Bohr emphasized the conditions under which physical concepts acquire meaning. According to Bohr, quantum phenomena cannot be meaningfully discussed without reference to the experimental arrangements that define the context of observation. Concepts such as position, momentum, or path are not attributes that can be simultaneously ascribed to a system in isolation; they are relational notions that depend on how the system is probed (Bohr, 1949). This relational view is formalized in modern quantum measurement theory, which treats measurement outcomes as context-dependent and mutually exclusive within a single experimental setup (Busch et al., 1996). Moreover, the necessity of contextuality finds a rigorous mathematical expression in the Kochen–Specker theorem, which demonstrates the impossibility of assigning non-contextual definite values to all quantum observables simultaneously (Kochen & Specker, 1967).

In this view, complementarity expresses a limitation on description, not a metaphysical claim about what exists. Bohr maintained that quantum mechanics is complete in the sense that it exhausts what can be consistently said about phenomena within the constraints imposed by experimental contexts. The mutual exclusivity of complementary descriptions reflects the impossibility of simultaneously realizing the experimental conditions required to define them, rather than a contradiction in the underlying physical process. Thus, Bohr's position was epistemological in character, emphasizing the inseparability of object and measuring arrangement in the articulation of physical knowledge.

### 3.3. What experiments decide—and what they leave open

Subsequent experimental developments, including violations of Bell inequalities and modern realizations of complementarity relations, are often interpreted as decisive in favor of Bohr's viewpoint. While such experiments undoubtedly rule out certain classes of local hidden-variable theories, their implications for the broader question of realism remain more nuanced. Experimental results constrain the structure of correlations and the viability of specific theoretical models, but they do not by themselves dictate a unique ontological interpretation.

In particular, demonstrations of the trade-off between path distinguishability and fringe visibility establish with great clarity the operational limits imposed by quantum mechanics (Englert, 1996; Durr et al., 1998). They show that no experimental arrangement can simultaneously yield maximal values of both quantities. However, this result pertains to what can be jointly accessed or represented within a given experimental context. It does not entail that the quantum system lacks an underlying state from which these mutually exclusive descriptions derive, nor does it resolve debates about whether such a state should be regarded as physically real, informational, or relational (Howard, 1994).

Importantly, these experiments do not favor Bohr's epistemology over Einstein's realism in any straightforward sense. Rather, they validate the structural predictions of quantum theory while remaining neutral on interpretive questions about what these structures represent at the ontological level. This distinction between operational confirmation and ontological commitment is crucial for understanding what the experimental record actually settles. This nuanced stance aligns with the concept of "contextual objectivity," which holds that quantum phenomena are objectively real but only within a well-defined measurement context (Grangier, 2002).

### 3.4. Epistemic complementarity: A reconciliatory framework

When viewed through the lens of modern measurement theory and decoherence, complementarity is most naturally interpreted as an epistemic principle. It specifies constraints on the simultaneous availability of certain kinds of information, given the structure of quantum interactions and the necessity of coupling systems to measurement devices and environments. These constraints are objective in the sense that they follow from the formalism of quantum mechanics and are borne out experimentally, yet they pertain to the conditions of knowledge acquisition rather than to the ultimate constitution of reality (Zurek, 2003; Schlosshauer, 2007).

This epistemic stance is further developed in the Quantum Bayesian (QBist) interpretation, which treats quantum states as representations of an agent's beliefs and complementarity as a constraint on how different beliefs can be simultaneously held (Fuchs & Schack, 2013).

This epistemic reading preserves the central insights of Bohr's position while leaving open the core concern articulated by Einstein. It acknowledges that quantum mechanics places unavoidable limits on what can be said about physical systems within a single descriptive framework, without concluding that reality itself is fragmented or observer-dependent in a strong ontological sense. Complementarity, understood this way, delineates the boundary between mutually incompatible modes of access to information, rather than serving as a verdict on the existence of unmeasured properties.

The resulting perspective reframes the Einstein–Bohr debate not as a contest with a definitive winner, but as an enduring tension between different criteria for theoretical adequacy. Einstein's insistence on descriptive completeness articulates a concern about whether a theory exhausts the physical content of reality, while Bohr's emphasis on contextuality addresses the conditions under which physical descriptions acquire meaning. Both concerns remain valid, but they operate at different levels: one pertaining to what a theory should describe, the other to how descriptions acquire content within the theory.

This reframing sets the stage for extending the discussion beyond quantum foundations. If complementarity captures a general structural feature of how information is accessed under certain constraints, similar limitations may arise in other domains where representation, context, and commitment interact. The following sections explore this possibility by examining whether analogous forms of complementarity operate in cognitive processes, thereby motivating the introduction of cognitive complementarity as a principled extension of the epistemic lessons drawn from quantum mechanics.

## 4. Cognitive Constraints and Non-Classical Uncertainty

Having established complementarity as an epistemic principle governing quantum measurement, we now turn to the domain of cognition to explore whether analogous constraints shape human reasoning under uncertainty. Classical models of judgment and decision-making, grounded in Boolean logic and Kolmogorovian probability, often fail to capture systematic patterns of real-world reasoning where context, order, and perspective play constitutive roles. This section examines these limitations, introduces alternative frameworks that accommodate non-classical uncertainty, and identifies structural parallels with quantum measurement that motivate a principle of cognitive complementarity.

### 4.1. When classical models of reasoning break down

Classical models of human reasoning have traditionally been grounded in Boolean logic and Kolmogorovian probability theory. Within this framework, beliefs are assumed to possess definite truth values, probabilities are required to obey the law of total probability, and decision-making is modeled as the optimization of expected utility over a fixed state space (Simon, 1955). Such models have proven effective in well-defined, low-uncertainty environments, and they continue to serve as normative benchmarks in economics and decision theory.

However, a substantial body of empirical work in psychology and cognitive science has demonstrated systematic and reproducible deviations from these classical assumptions. Phenomena such as order effects, context dependence, preference reversals, and violations of the sure-thing principle indicate that human judgments often depend on how questions are framed and sequenced (Tversky & Kahneman, 1981; Busemeyer & Bruza, 2012). These deviations are not merely artifacts of noise or irrationality; rather, they reflect structural features of cognition operating under uncertainty, limited information, and contextual constraints. Classical models, while mathematically consistent, frequently lack the expressive resources needed to capture these features without introducing ad hoc modifications.

## 4.2. Contextuality and non-Boolean structures in cognition

In response to these limitations, alternative formal approaches have been developed that relax classical assumptions about probability and logic. Of particular relevance are models that treat cognitive states as inherently contextual, such that the outcome of a judgment depends not only on prior beliefs but also on the specific conditions under which information is elicited (Wang et al., 2013). In these approaches, the act of questioning or decision-making plays a constitutive role, shaping the cognitive state rather than merely revealing a pre-existing value.

Research in quantum cognition has shown that the mathematical structures of quantum probability—Hilbert spaces, superposition, and non-commuting observables—can successfully model a range of cognitive phenomena that resist classical explanation (Busemeyer & Bruza, 2012; Pothos & Busemeyer, 2009). Importantly, these models do not imply that cognitive processes are physically quantum or that the brain operates as a quantum computer. Instead, they employ quantum formalisms as generalized probabilistic frameworks capable of representing contextuality, interference effects, and order dependence in decision-making. The success of such models suggests that cognition, like quantum measurement, may involve structural constraints that preclude the simultaneous applicability of all descriptive perspectives.

These structural parallels find preliminary support in existing cognitive research. For instance, dual-process theory (Kahneman, 2011) distinguishes intuitive (System 1) and analytical (System 2) thinking, but also acknowledges their mutual interference under cognitive load. Similarly, predictive processing models (Clark, 2013) propose that the brain balances precision-weighted prediction errors across hierarchical models—a trade-off between top-down coherence and bottom-up sensitivity. Metacognitive monitoring (Nelson & Narens, 1990) further illustrates how agents regulate belief commitment based on contextual uncertainty. These examples suggest that cognitive systems routinely navigate representational trade-offs analogous to complementarity, even if not previously framed in epistemic-structural terms.

### 4.3. Decision commitment and representational reduction

A central feature shared by many non-classical cognitive models is their recognition that decisions involve a form of representational reduction. Prior to commitment, an agent may entertain multiple, mutually incompatible possibilities—competing interpretations, intentions, or evaluations. The act of deciding selects one such possibility, rendering others inactive or inaccessible for the purpose of subsequent reasoning (Atmanspacher et al., 2002). This transition from a plural or indeterminate cognitive state to a determinate outcome is not merely a passive update; it often reshapes the agent's future judgments and available options.

Empirical studies indicate that the timing and context of decision commitment can significantly influence outcomes. Premature commitment may simplify reasoning but at the cost of flexibility and sensitivity to new information, while delayed commitment can preserve optionality at the expense of increased cognitive load (Gollwitzer, 1990). These trade-offs suggest that decision-making is constrained by structural limits on how information can be represented and maintained simultaneously, rather than by a simple lack of computational capacity. The process mirrors, in a cognitive register, the measurement-induced collapse of quantum superposition—a parallel we now explore explicitly.

### 4.4. Structural parallels with quantum measurement

The limitations observed in cognitive reasoning under uncertainty bear a structural resemblance to the constraints imposed by quantum complementarity. In both cases, attempts to extract precise, determinate information along one dimension can suppress access to other relevant features. In quantum measurement, increased path distinguishability reduces fringe visibility (Englert, 1996); in cognition, increased commitment to a particular representation can reduce sensitivity to alternative interpretations or contextual cues (Busemeyer & Bruza, 2012). While these parallels do not license a direct physical analogy, they motivate a comparative analysis at the level of information and representation.

Crucially, these constraints are not merely subjective or idiosyncratic. Just as complementarity reflects objective features of quantum interactions, cognitive limitations arise from systematic features of how information is processed, represented, and acted upon by finite agents. The existence of such limitations challenges the assumption that all relevant aspects of a situation can be simultaneously optimized or represented within a single cognitive framework.

### 4.5. Toward cognitive complementarity

Taken together, these observations suggest the need for a principled account of how incompatible cognitive representations coexist and how transitions between them are managed. If reasoning under non-classical uncertainty involves trade-offs analogous to those encountered in quantum measurement, then a framework that explicitly acknowledges such trade-offs may offer a more accurate description of cognitive behavior.

This motivates the introduction of cognitive complementarity as a structural principle governing reasoning in contexts where information is contextual, incomplete, or mutually constraining.

Cognitive complementarity, as developed in the next section, is not proposed as a metaphorical borrowing from physics, but as a formal and conceptual extension of the epistemic lessons derived from quantum measurement theory. It provides a framework for understanding why certain cognitive perspectives cannot be simultaneously optimized, and how agents can nonetheless navigate such constraints without premature representational collapse. By articulating how incompatible cognitive representations are constrained and coordinated, this framework prepares the ground for defining quantum intuition as a cognitive capacity adapted to operating within such limits.

## 5. Cognitive Complementarity

Building on the structural parallels between quantum measurement and cognitive reasoning under uncertainty, this section introduces the principle of cognitive complementarity. We define this principle, illustrate its operation through practical examples of incompatible cognitive representations, examine the process and costs of cognitive collapse, and situate cognitive complementarity as an epistemic constraint analogous to—yet distinct from—its quantum counterpart.

### 5.1. Defining cognitive complementarity

Cognitive complementarity is introduced here as a structural principle governing reasoning in contexts where information is inherently contextual, incomplete, or mutually constraining. It refers to the limitation that certain cognitive representations, while individually well-defined and operationally useful, cannot be simultaneously maintained or optimized without loss of coherence or relevance. This limitation is not reducible to finite computational resources or attentional capacity alone; rather, it reflects deeper constraints on how information is represented, integrated, and acted upon by cognitive agents (Atmanspacher et al., 2002; Busemeyer & Bruza, 2012).

Under cognitive complementarity, different representational modes correspond to different questions posed to a situation. Each mode enables access to specific features while suppressing others. No single representation exhausts all relevant aspects simultaneously, and attempts to force such simultaneity result in instability, inconsistency, or premature closure. Cognitive complementarity thus characterizes a class of reasoning problems in which trade-offs between incompatible but informative perspectives are unavoidable.

### 5.2. Incompatible cognitive representations in practice

In practical reasoning, agents routinely navigate between representations that emphasize different dimensions of a problem. For example, analytic precision may require abstraction and simplification, whereas holistic understanding may depend on preserving contextual relations and ambiguity (Kahneman, 2011). Similarly, determinacy supports decisive action, while openness preserves adaptability to changing conditions.

Cognitive complementarity arises when such representational modes cannot be jointly maximized, even in principle, because the conditions that stabilize one mode undermine the other. These incompatibilities are not pathological. On the contrary, they often reflect the richness of the underlying situation. Cognitive complementarity becomes salient precisely in domains where uncertainty is not merely epistemic noise to be reduced, but a structural feature of the problem space. In such domains, insisting on a single, globally optimal representation can obscure relevant information and degrade performance.

### 5.3. Cognitive collapse and its regulatory costs

A central process governed by cognitive complementarity is decision commitment, understood as the transition from a state of representational plurality to a determinate course of action or judgment. Prior to commitment, multiple representations may coexist, each partially supported by available information. Commitment selects one representation as operative, effectively suppressing alternatives for the purposes of action and subsequent reasoning.

This transition can be described as a form of cognitive collapse, not in a metaphysical sense, but as an operational reduction of representational degrees of freedom (Pothos & Busemeyer, 2009). Cognitive collapse is necessary for action, yet it carries costs. Once a representation is stabilized, sensitivity to alternative framings diminishes, and subsequent information is often assimilated in a biased or constrained manner. The timing and context of collapse therefore play a critical role in determining cognitive flexibility and effectiveness.

Cognitive complementarity highlights a fundamental trade-off in reasoning under uncertainty. Premature collapse—commitment before sufficient contextual information has been integrated—can lead to overconfidence, rigidity, and systematic error. Conversely, excessively delayed collapse may preserve flexibility but at the cost of indecision, cognitive overload, or missed opportunities for timely action (Gollwitzer, 1990). Neither extreme is universally optimal.

Effective reasoning thus requires navigating between these poles, maintaining representational plurality long enough to capture relevant contextual structure, while committing decisively when conditions warrant. Cognitive complementarity provides a framework for understanding why this balance cannot be reduced to a single optimal strategy independent of context.

### 5.4. An epistemic constraint, not a metaphysical claim

Importantly, cognitive complementarity should be understood as an epistemic constraint, not as a claim about the metaphysical nature of mind. It delineates limits on what can be simultaneously represented and utilized within a single cognitive episode, given the structure of information and the requirements of action. These limits are objective in the sense that they arise systematically across agents and tasks, yet they pertain to modes of access and representation rather than to the existence of underlying states (Howard, 1994).

By articulating these constraints explicitly, cognitive complementarity provides a principled bridge between the epistemic lessons of quantum measurement theory and the study of reasoning under uncertainty. It establishes the conceptual space within which a specific cognitive capacity—capable of operating effectively under such constraints—can be defined. This epistemic constraint can be formalized in a minimal model, presented in Box 1.

---

**Box1: A Minimal Model of Cognitive Complementarity**

Let an agent's belief state about a decision be represented by a vector in a 2D Hilbert space:

$$|\psi\rangle = \alpha |A\rangle + \beta |B\rangle$$

where |A⟩ and |B⟩ correspond to two incompatible interpretations (e.g., "risk" vs. "opportunity").

Cognitive measurement corresponds to a projection onto one basis ($e.g., \{|A\rangle, |A^\perp\rangle\}$)

Complementarity constraint: The certainty in one basis reduces certainty in the other, formalized as:

$$\Delta_A^2 + \Delta_B^2 \geq C$$

Where $\Delta_X^2$ is the variance in judgment along basis $X$, and $C$ is a cognitive-complementarity constant.

---

The following section introduces quantum intuition as such a capacity. Rather than proposing a new metaphysical faculty, quantum intuition is presented as a mode of cognition adapted to managing epistemic complementarity: sustaining representational plurality, delaying collapse when appropriate, and resolving incompatibilities in a context-sensitive manner.

# 6. Quantum Intuition

Within the framework established by cognitive complementarity, this section introduces quantum intuition as a cognitive capacity adapted to reasoning under epistemic constraints that preclude the simultaneous optimization of incompatible representations. Rather than a mysterious or metaphysical faculty, quantum intuition is presented as a testable mode of cognition that enables agents to sustain representational plurality, regulate commitment, and navigate uncertainty in a context-sensitive manner.

### 6.1. What is quantum intuition?

Quantum intuition refers to an agent's ability to operate coherently in states of representational plurality, sustaining multiple, mutually constraining perspectives without premature commitment, and to resolve these constraints in a context-sensitive manner when action is required. Quantum intuition does not imply access to hidden variables, nonlocal insight, or any privileged epistemic status. Instead, it characterizes a mode of cognition in which uncertainty is treated as structurally informative rather than as a deficiency to be eliminated as quickly as possible (Busemeyer & Bruza, 2012; Atmanspacher et al., 2002).

The defining feature of quantum intuition is not the absence of commitment, but the timing and manner of commitment under conditions where no single representation can be regarded as complete. It reflects a capacity to manage epistemic tension productively—to delay cognitive collapse when context is still unfolding, and to commit decisively when sufficient structure has emerged.

### 6.2. Beyond classical intuition

In classical psychological accounts, intuition is often associated with rapid, heuristic-based judgments that operate below the level of explicit deliberation (Kahneman, 2011). Such intuitions are typically shaped by pattern recognition, past experience, and associative processing, and they are often contrasted with slower, analytical reasoning. While this distinction captures important features of human cognition, it does not adequately address situations in which the structure of the problem itself resists classical representation.

Quantum intuition differs from classical intuition in three key respects. First, it is context-sensitive rather than context-agnostic: its effectiveness depends on the agent's capacity to track how changes in framing alter the space of relevant possibilities. Second, it is plural rather than singular, sustaining incompatible representations instead of collapsing immediately to a single dominant heuristic. Third, it is reflexively constrained, recognizing that any act of commitment reshapes the cognitive landscape and limits subsequent access to alternative perspectives.

### 6.3. Sustaining epistemic superposition

A useful way to characterize quantum intuition is through its capacity to sustain what we term epistemic superposition. Prior to commitment, an agent guided by quantum intuition maintains multiple candidate representations as simultaneously operative, without forcing an artificial resolution. This state is not one of indecision or confusion, but of structured openness, in which the relations among competing representations are actively monitored (Pothos & Busemeyer, 2009).

Epistemic superposition is inherently unstable and cannot be maintained indefinitely. Cognitive resources are finite, and action eventually requires reduction to a determinate course. Quantum intuition therefore does not eliminate cognitive collapse; rather, it regulates collapse, delaying it when premature commitment would obscure critical contextual information, and enabling it when sufficient structure has emerged to justify resolution.

### 6.4. Formal and operational foundations

Although quantum intuition is introduced here at a conceptual level, it is amenable to formal and empirical treatment. From a modeling perspective, quantum probability frameworks provide one possible formal language for representing epistemic superposition, contextuality, and non-commuting cognitive observables (Busemeyer & Bruza, 2012; Wang et al., 2013). Within such models, belief states can be represented as vectors or density operators, and decision commitment as a projection conditioned on context. Importantly, this formalism is employed as a representational tool rather than as a claim about the physical substrate of cognition.

Operationally, quantum intuition may be investigated through tasks that manipulate contextual framing, information order, and commitment timing. Behavioral signatures of quantum intuition would include enhanced performance in environments characterized by deep uncertainty, reduced susceptibility to framing-induced bias under delayed commitment, and adaptive switching between representational modes. These signatures distinguish quantum intuition from both rigid analytical strategies and purely heuristic responses.

### 6.5. A non-mystical, testable construct

It is essential to emphasize that quantum intuition is neither a metaphysical postulate nor a claim about consciousness exerting causal influence on physical systems. It does not assert that cognition collapses quantum states, nor that mental processes exploit physical quantum coherence. Instead, it identifies a cognitive strategy shaped by the same epistemic constraints that govern quantum measurement: constraints on what can be jointly accessed, represented, and acted upon.

By formulating quantum intuition in this manner, the concept remains fully compatible with naturalistic accounts of cognition and open to empirical validation. Its value lies not in explanatory reach alone, but in its capacity to generate testable hypotheses about reasoning, learning, and decision-making under non-classical uncertainty (Simon, 1955; Gollwitzer, 1990). The following section examines these implications explicitly, outlining how quantum intuition may inform experimental design, behavioral prediction, and the study of reasoning in complex, context-dependent environments.

## 7. Implications and Testability

The introduction of quantum intuition as a cognitive capacity raises important questions about how reasoning under uncertainty is studied, modeled, and improved. This section explores the implications of the framework for real-world decision contexts, derives testable behavioral predictions, proposes experimental paradigms for investigation, and considers the role of quantum intuition in learning and adaptation.

### 7.1. Reasoning under deep uncertainty

In many real-world contexts—scientific inquiry, strategic planning, policy design, and complex learning—agents must act without access to complete or simultaneously consistent information (Simon, 1955). Classical normative models typically assume that uncertainty can be reduced through accumulation of data or optimization over fixed probability spaces. The framework developed here suggests that, in certain domains, uncertainty is not merely a temporary epistemic deficit but a structural feature of the problem itself (Busemeyer & Bruza, 2012).

Within such domains, performance may depend less on rapid convergence to a single representation than on the ability to maintain representational plurality long enough for relevant contextual structure to emerge. Quantum intuition provides a principled account of how such plurality can be sustained without degenerating into indecision. It therefore offers an explanatory lens for why strategies emphasizing delayed commitment, flexible reframing, and sensitivity to context often outperform rigid optimization in complex environments (Kahneman, 2011).

Beyond the laboratory, this framework has implications for education (teaching students to tolerate ambiguity and delay closure in complex problem-solving), AI design (building systems that maintain multiple hypotheses in uncertain environments), and organizational decision-making (avoiding premature consensus in strategic planning). In each case, the goal is not to avoid commitment indefinitely, but to optimize its timing relative to contextual richness.

### 7.2. Behavioral predictions and experimental paradigms

If quantum intuition captures a genuine cognitive capacity, it should give rise to identifiable behavioral signatures. Several testable predictions follow from the framework developed in this paper:

I. Contextual resilience: Agents exhibiting higher quantum intuition should show reduced susceptibility to framing effects when allowed to delay commitment, compared to agents who commit early (Tversky & Kahneman, 1981).

II. Adaptive switching: Such agents should demonstrate greater facility in switching between incompatible representations as task demands change, without incurring excessive cognitive cost.

III. Performance under non-commuting cues: In tasks where outcomes depend on order-dependent evidence or conflicting signals, agents employing quantum-intuitive strategies should outperform both purely heuristic and purely analytical approaches (Pothos & Busemeyer, 2009).

These predictions distinguish quantum intuition from general intelligence or processing speed and situate it as a specific response to epistemic constraints.

The framework admits several experimental implementations. Behavioral tasks can be designed in which participants receive information sequentially under varying contextual framings, with the opportunity to commit at different stages. By manipulating the cost of premature versus delayed commitment, such tasks can probe the trade-offs central to cognitive complementarity. Measures of interest would include decision timing, outcome quality, sensitivity to order effects, and post-decision adaptability.

A candidate experimental paradigm is the Dynamic Framing Task, in which participants evaluate a scenario (e.g., a business decision) under two sequentially presented frames (e.g., "growth opportunity" vs. "potential loss"). The key manipulation is the timing of commitment: early commitment (after first frame) vs. delayed commitment (after both frames). Measures include decision quality, confidence, and post-decision flexibility. We predict that high quantum-intuition participants will outperform in delayed conditions, showing less framing bias and better integration of conflicting cues. This design directly tests the trade-off between premature collapse and sustained epistemic superposition.

### 7.3. Formal modeling and neurocognitive correlates

Computational modeling provides a complementary avenue for investigation. Quantum probability models, already employed in quantum cognition research, can be used to formalize epistemic superposition and representational collapse in decision-making tasks (Busemeyer & Bruza, 2012; Wang et al., 2013). Comparing the predictive performance of these models against classical probabilistic models across different task regimes would offer quantitative tests of the framework.

Neurocognitive investigations may also prove informative. While the present framework makes no claims about physical quantum processes in the brain, it predicts differential neural signatures associated with sustained representational plurality versus early commitment. These may include patterns related to cognitive control, conflict monitoring, and attentional flexibility, which can be examined using established neuroimaging and electrophysiological methods.

### 7.4. Learning, adaptation, and scientific practice

Beyond isolated decision tasks, quantum intuition has implications for learning and adaptation over extended timescales. Learning in environments characterized by contextual instability or competing explanatory frameworks may benefit from strategies that resist early convergence (Gollwitzer, 1990). Agents capable of sustaining epistemic superposition may explore hypothesis spaces more effectively and avoid local optima imposed by premature model fixation.

This perspective aligns with observations in scientific practice, where major advances often arise from prolonged tolerance of conceptual tension rather than rapid resolution (Kuhn, 1962). Quantum intuition thus provides a cognitive account of why such tolerance may be adaptive, grounding it in structural constraints rather than individual temperament or stylistic preference.

### 7.5. Scope and limitations

Finally, it is important to delimit the scope of the present framework. Quantum intuition is not proposed as a universal solution to all cognitive problems, nor as a replacement for analytical reasoning. Its relevance is confined to contexts where information is contextual, incompatible, or irreducibly incomplete. In well-defined, stable environments, classical optimization and rapid commitment may remain superior.

This scope limitation also helps preempt potential critiques. First, one might argue that quantum intuition merely relabels existing constructs like cognitive flexibility or metacognition. However, quantum intuition specifically addresses the structural incompatibility of representations and the epistemic timing of commitment—dimensions not fully captured by flexibility alone. Second, the analogy with quantum mechanics could be seen as superficial. Here, it is crucial to reiterate that the analogy is structural and epistemic, not ontological; we import constraints on representation, not physical mechanisms. Third, the utility of quantum probability models may be questioned. While such models offer a natural formalism for representing superposition and contextuality, the core of the framework is the complementarity principle itself, which can also be explored using classical models with context-dependent constraints.

By articulating these boundaries explicitly, the framework avoids overgeneralization and invites empirical refinement. The testability of quantum intuition—its susceptibility to confirmation, modification, or rejection—constitutes a central strength of the present approach. The concluding section synthesizes these implications, returning to the Einstein-Bohr debate to reflect on how epistemic limits in physics can illuminate parallel constraints in cognition without collapsing one domain into the other.

## 8. Discussion & Conclusion

This article has examined the persistent tension between epistemic access and ontological commitment in quantum mechanics through the lens of the Einstein–Bohr debate, and extended these considerations into the cognitive domain. By reframing complementarity as an epistemic principle, introducing cognitive complementarity, and proposing quantum intuition as a testable capacity, we have sought to bridge foundational physics and cognitive science in a principled, non-metaphorical manner. This concluding section synthesizes the core arguments, reflects on their implications, and outlines open questions for future research.

### 8.1. Complementarity as an epistemic limit

The preceding analysis supports a reading of quantum complementarity as a constraint on what can be simultaneously accessed, represented, and operationally described, rather than as a definitive statement about the ontological structure of reality. Modern formulations based on path distinguishability, fringe visibility, information flow, and decoherence clarify the physical origin of complementarity without requiring the abandonment of realist intuitions (Englert, 1996; Zurek, 2003). The loss of interference arises from the redistribution of information through correlations, not from the annihilation of underlying physical possibilities.

Interpreting complementarity in this epistemic sense preserves the empirical content of quantum mechanics while maintaining conceptual openness regarding questions of completeness and realism. This distinction is particularly important in light of contemporary experiments, whose increasing precision sharpens operational constraints but does not, by itself, adjudicate ontological debates (Fine, 1986). The present work therefore aligns with a cautious interpretive stance: experimental confirmation of complementarity delineates the structure of measurement outcomes and representations, not the full scope of what may exist independently of those representations.

### 8.2. Reframing the Einstein–Bohr debate

Within this framework, the historical Einstein–Bohr debate can be understood not as a conflict with a definitive resolution, but as an enduring tension between two criteria for theoretical adequacy. Einstein's insistence on descriptive completeness articulated a concern about whether a theory exhausts the physical content of reality, while Bohr's emphasis on contextuality addressed the conditions under which physical descriptions acquire meaning (Bohr, 1949; Einstein et al., 1935). The epistemic reading of complementarity developed here accommodates both perspectives without forcing a reductive synthesis.

Rather than viewing subsequent experimental developments as vindicating one position at the expense of the other, the analysis suggests that these developments clarify the domain of validity of each stance. Bohr's conclusions regarding the limits of description are strongly supported at the operational level, while Einstein's concern about the distinction between knowledge and reality remains conceptually coherent. This reframing shifts the debate from questions of victory or defeat to questions of scope and interpretation.

### 8.3. Cognitive complementarity and quantum intuition: A principled extension

The extension of complementarity into the cognitive domain is motivated by this same distinction. If quantum measurement reveals structural limits on representation imposed by the conditions of access to information, it is plausible that analogous limits arise in cognitive systems confronted with contextual, incompatible, or irreducibly uncertain information (Busemeyer & Bruza, 2012). Cognitive complementarity captures this possibility without importing physical assumptions about the substrate of cognition.

Within this framework, quantum intuition emerges as a specific cognitive capacity adapted to managing epistemic complementarity. Its defining features—sustaining representational plurality, regulating the timing of commitment, and resolving incompatible perspectives in a context-sensitive manner—are directly responsive to the constraints identified in both quantum measurement and cognitive reasoning under uncertainty. By formulating quantum intuition as a testable construct rather than a metaphysical postulate, the present work seeks to avoid two common pitfalls: treating intuition as an ineffable faculty beyond empirical scrutiny, and reducing all non-classical reasoning to error or irrationality.

### 8.4. Open questions and interdisciplinary horizons

Several limitations and open questions remain. First, the concept of quantum intuition requires further operationalization, and its relationship to existing constructs in cognitive science (such as cognitive flexibility, tolerance of ambiguity, or metacognitive monitoring) demands systematic empirical investigation. Second, while quantum probability models provide a promising formal language, their scope and limits in capturing real-world reasoning must be carefully assessed against alternative frameworks (Wang et al., 2013). Finally, the analogy between physical and cognitive complementarity, while structurally motivated, should not be overextended. Not all cognitive constraints mirror quantum constraints, and not all features of quantum measurement have meaningful cognitive counterparts.

Nevertheless, the broader implication of this analysis is methodological. Quantum mechanics has long challenged classical intuitions about measurement and description, leading to interpretive debates that often oscillate between ontological speculation and operational minimalism. By extracting the epistemic lessons of complementarity and applying them to cognition, the present framework illustrates how foundational insights from physics can inform other domains without collapsing disciplinary boundaries. Conversely, the cognitive perspective feeds back into foundational discussions by highlighting the role of representational constraints and commitment in shaping what counts as an acceptable description.

Future work should focus on empirical tests of quantum intuition, refinement of the cognitive complementarity framework, and exploration of its applications in education, decision training, and the design of adaptive systems. In doing so, we may find that the enduring challenges posed by quantum mechanics continue to offer valuable insights—not only into the nature of physical measurement, but into the limits and possibilities of human understanding itself.

Looking forward, we propose a three-part research agenda: (1) empirical validation of quantum intuition through the Dynamic Framing Task and neuroimaging studies of representational maintenance; (2) theoretical extension of cognitive complementarity to collective and cultural reasoning; and (3) applied development of training protocols to enhance quantum intuition in professionals operating under deep uncertainty. By pursuing this agenda, we may not only clarify the epistemic structure of reasoning but also cultivate cognitive capacities suited to an increasingly complex world.